\documentclass[prl,twocolumn,floatix,amssymb,amsmath,superscriptaddress]{revtex4-1}
\usepackage{graphicx}
\usepackage{epsfig,color}
\usepackage{bbm,amsmath,amssymb}
\begin{document}

\newcommand{\ket}[1]{\ensuremath{\left|{#1}\right\rangle}}
\newcommand{\bra}[1]{\ensuremath{\left\langle{#1}\right|}}
\newcommand{\quadr}[1]{\ensuremath{{\not}{#1}}}
\newcommand{\quadrd}[0]{\ensuremath{{\not}{\partial}}}
\newcommand{\slpar}{\partial\!\!\!/}
\newcommand{\gtrescero}{\gamma_{(3)}^0}
\newcommand{\gtresuno}{\gamma_{(3)}^1}
\newcommand{\gtresi}{\gamma_{(3)}^i}

\title{Quantum Simulation of the Majorana Equation and Unphysical Operations}
\author{J. Casanova}
\affiliation{Departamento de Qu\'{\i}mica F\'{\i}sica, Universidad del Pa\'{\i}s Vasco -- Euskal Herriko Unibertsitatea, Apdo.\ 644, 48080 Bilbao, Spain}
\author{C. Sab{\'i}n}
\affiliation{Instituto de F\'{\i}sica Fundamental, CSIC, Serrano 113-bis, 28006 Madrid, Spain}
\author{J. Le{\'o}n}
\affiliation{Instituto de F\'{\i}sica Fundamental, CSIC, Serrano 113-bis, 28006 Madrid, Spain}
\author{I. L. Egusquiza}
\affiliation{Departamento de F\'{\i}sica Te\'{o}rica, Universidad del Pa\'{\i}s Vasco -- Euskal Herriko Unibertsitatea, Apdo.\ 644, 48080 Bilbao, Spain}
\author{R.~Gerritsma}
\affiliation{Institut f\"ur Quantenoptik und Quanteninformation, \"Osterreichische Akademie der Wissenschaften, Otto-Hittmair-Platz 1, A-6020 Innsbruck, Austria}
\affiliation{Institut f\"ur Experimentalphysik, Universit\"at Innsbruck, Technikerstrasse 25, A-6020 Innsbruck, Austria}
\author{C.~F. Roos}
\affiliation{Institut f\"ur Quantenoptik und Quanteninformation, \"Osterreichische Akademie der Wissenschaften, Otto-Hittmair-Platz 1, A-6020 Innsbruck, Austria}
\affiliation{Institut f\"ur Experimentalphysik, Universit\"at Innsbruck, Technikerstrasse 25, A-6020 Innsbruck, Austria}
\author{J.~J. Garc{\'i}a-Ripoll}
\affiliation{Instituto de F\'{\i}sica Fundamental, CSIC, Serrano 113-bis, 28006 Madrid, Spain}
\author{E. Solano}
\affiliation{Departamento de Qu\'{\i}mica F\'{\i}sica, Universidad del Pa\'{\i}s Vasco -- Euskal Herriko Unibertsitatea, Apdo.\ 644, 48080 Bilbao, Spain}
\affiliation{IKERBASQUE, Basque Foundation for Science, Alameda Urquijo 36, 48011 Bilbao, Spain}

\begin{abstract}
We design a quantum simulator for the Majorana equation, a non-Hamiltonian relativistic wave equation that might describe neutrinos and other exotic particles beyond the standard model. The simulation demands the implementation of charge conjugation, an unphysical operation that opens a new front in quantum simulations, including other discrete symmetries as complex conjugation and time reversal. Furthermore, we describe how to implement this general method in trapped ions.
\end{abstract}

\maketitle

A quantum simulator is a device engineered to reproduce the properties of an ideal quantum model. This still-emerging topical area has generated a remarkable exchange of scientific knowledge between apparently unconnected subfields of physics. In terms of applications, it allows for the study of quantum systems that cannot be efficiently simulated on classical computers~\cite{Feynman:1982}. While a quantum computer would also implement a universal quantum simulator~\cite{Lloyd:1996}, only particular systems have been simulated up to now using dedicated quantum simulators~\cite{Buluta:2009}. Still, there is a wealth of successful cases, such as spin models~\cite{Friedenauer:2008,Kim:2010}, quantum chemistry~\cite{Lanyon:2010} and  quantum phase transitions~\cite{Greiner:2002}. The quantum simulation of fermionic systems~\cite{Casanovafermions} and relativistic quantum physics have also attracted recent attention, reproducing dynamics and effects currently out of experimental reach. Examples include black holes in Bose-Einstein condensates~\cite{Lahav:2010}, quantum field theories~\cite{Casanovaqft, Preskill2011}  and recent quantum simulations of relativistic quantum effects as {\it Zitterbewegung}, Klein paradox and interacting relativistic particles~\cite{Lamata:2007,Gerritsma:2010,Casanova:2010,Gerritsma:2011, LamataNJP} in trapped ions.

In this paper, we show how the Majorana equation~\cite{Majorana:1937} can be simulated in an analog quantum simulator, having as a key requirement  the implementation of complex conjugation of the wavefunction. In this manner, we are able to propose this and other unphysical operations such as charge conjugation and time reversal, constituting a novel toolbox of accessible quantum operations in the general frame of quantum simulations. While quantum simulators may soon realize calculations impossible for classical computers, we show here the possibility of implementing quantum dynamics that are impossible for our quantum world.

The Majorana equation is a relativistic wave equation for fermions where the mass term contains the charge conjugate of the spinor, $\psi_{c}$,
\begin{equation}
\label{majorana}
i \hbar \slpar \psi = m c \psi_{c} .
\end{equation}
Here, $\slpar = \gamma^{\mu} \partial_{\mu}$ and $\gamma_{\mu}$ are the Dirac matrices~\cite{Thaller:1992}, while the non-Hamiltonian character stems from the simultaneous presence of  $\psi$ and $\psi_{c}$. The significance of the Majorana equation rests on the fact that it can be derived from first principles in a similar fashion as the Dirac equation~\cite{Majorana:1937,Zee:2003}. Both wave equations are Lorentz invariant but the former preserves helicity and does not admit stationary solutions. The Majorana equation is considered a possible model~\cite{Aste:2010} for describing exotic particles in supersymmetric theories --photinos and gluinos--, or in grand unified theories, as is the case of neutrinos. Indeed, the discussion of whether neutrinos are Dirac or Majorana particles still remains open~\cite{Giunti:2007}. Nevertheless, despite the similar naming, this work is neither related to the Majorana fermions (modes) in many-body systems~\cite{Kitaev:2001, Mezzacapo2011}, nor to the Majorana fermions (spinors) in the Dirac equation~\cite{Aste:2010,Wilczek:2009}.

In order to simulate the Majorana equation, we have to solve a fundamental problem: the physical implementation of antilinear and antiunitary operations in a quantum simulator. Here, we introduce a mapping~\cite{Comment1} by which complex conjugation, an unphysical operation, becomes a unitary operation acting on an enlarged Hilbert space. The mapping works in arbitrary dimensions and can be immediately applied on different experimental setups. We show how to simulate the Majorana equation in 1+1 dimensions and other unphysical operations using two trapped ions. We also give a recipe for measuring observables and a roadmap towards more general scenarios. In this sense, this work provides a novel toolbox for quantum simulations.

\begin{figure*}[t]
\hspace*{-0.4cm}
\includegraphics[width=0.8\linewidth]{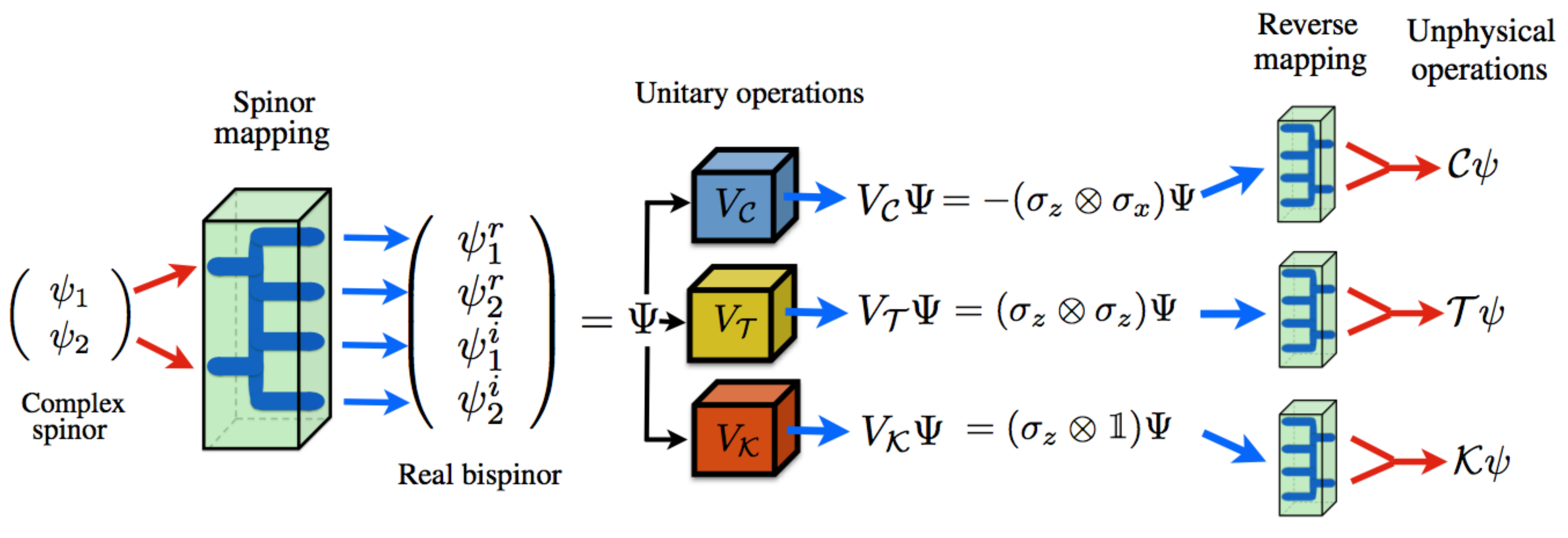}
  \caption{Diagram showing the different steps involved in the quantum simulation of unphysical operations in 1+1 dimensions.\vspace{-0.5cm}}
  \label{fig:mapping}
\end{figure*}

There are three discrete symmetries~\cite{Streater:2000} which are central to quantum mechanics and our understanding of particles, fields and their interactions: parity, $\mathcal{P}$, time reversal, $\mathcal{T}$, and charge conjugation, $\mathcal{C}$. None of these operations can be carried out in the real world: $\mathcal{P}$  involves a global change of physical space, while $\mathcal{C}$ and $\mathcal{T}$ are antiunitaries. However, there is no apparent fundamental restriction for implementing them in physical systems that simulate quantum mechanics. We will focus on the study of antiunitary operations, which can be decomposed into a product of a unitary, ${\mathcal{U}}_{\mathcal{C}}$ or ${\mathcal{U}}_{\mathcal{T}}$, and complex conjugation, $\mathcal{K}\psi=\psi^*$. We consider the mapping of the quantum states of an $n$-dimensional complex Hilbert space, $\mathbb{C}_n$, onto a real Hilbert space~\cite{Comment1}, $\mathbb{ R}_{2n}$,
\begin{equation}
 \psi \in \mathbb{C}_n \to \Psi = \frac{1}{2} \left(
    \begin{array}{c}\psi + \psi^*\\i(\psi^* - \psi)\end{array}
  \right) \in \mathbb{R}_{2n} .
\label{map}
\end{equation}
This mapping can be implemented by means of an auxiliary two-level system, such that $\mathbb{R}_{2n} \in \mathcal{H}_{2} \otimes \mathcal{H}_{n}$. In this manner, the complex conjugation of the simulated state becomes a local unitary $V_{\mathcal K}$ acting solely on the ancillary space, $\mathcal{K}\psi = \psi^*\to V_{\mathcal K} \Psi = (\sigma_z\otimes\openone)\Psi,$ and thus physically implementable for a wavefunction of arbitrary dimensions. Furthermore, unitaries and observables can also be mapped onto the real space, $O\to \Theta = \openone\otimes O_r -i \sigma^y\otimes O_i,$ where $O_{r}=\frac{1}{2}(O + \mathcal{K}O\mathcal{K})$ and  $O_{i}= - \frac{i}{2}(O - \mathcal{K}O\mathcal{K}),$ preserving unitarity and Hermiticity. The proposed simulator also accommodates the antiunitary operations $\mathcal{C} = {\mathcal U}_{\mathcal{C}}\mathcal{K}$ and $\mathcal{T} = {\mathcal U}_{\mathcal{T}}\mathcal{K}$. To this end, we have to choose a particular representation that fixes the unitaries ${\mathcal U}_{\mathcal{C}}$ and ${\mathcal U}_{\mathcal{T}}$, as will be shown below.

We possess now the basic tools to simulate the Majorana equation (\ref{majorana}). The expression for the charge conjugate spinor is given by $\psi_c = \mathcal{W}\gamma^0\mathcal{K} \psi ,$ with $\mathcal{W}$ a unitary matrix satisfying $\mathcal{W}^{-1}\gamma^\mu\mathcal{W}=-\left(\gamma^\mu\right)^T.$ We illustrate now the proposed quantum simulation with the case of 1+1 dimensions. Here, a suitable representation of charge conjugation is $\psi_c = i \sigma_y\sigma_z \psi^*$, that is ${\mathcal W} =  i \sigma_y$, and the Majorana equation reads
\begin{equation}
  \label{1+1}
  i\hbar\partial_{t}\psi = c\sigma_xp_{x}\psi - imc^2\sigma_{y} \psi^{*},
\end{equation}
where $p_{x}= -i\hbar\partial_{x}$ is the momentum operator. Note that Eq.~(\ref{1+1}) is not  Hamiltonian, ($i\hbar\partial_t \psi \neq H \psi$). This is due to the presence of a complex conjugate operation in the right-hand side of Eq.~(\ref{1+1}), which is not a linear Hermitian operator. Surprisingly, through our mapping~(\ref{map}),
\begin{equation}
\left(
\begin{array}{c}
\psi_1 \\ \psi_2
\end{array}
\right)
\in \mathbb{C}_2 \to \Psi = \left(
\begin{array}{c}
\psi_1^r \\ \psi_2^r \\ \psi_1^i \\ \psi_2^i
\end{array}
  \right) \in \mathbb{R}_{4} ,
\label{map 1+1}
\end{equation}
the Majorana equation for a complex spinor becomes a 3+1 Dirac equation with dimensional reduction, $p_y,p_z=0$, and a four-component real bispinor
\begin{equation}
\label{eq:majorspin31}
i\hbar \partial_{t}\Psi =
\left[c(\openone \otimes \sigma_x) p_x  - mc^{2} \sigma_x\otimes\sigma_y\right] \Psi .
\end{equation}
Here, the dynamics preserves the reality of the bispinor $\Psi$ and cannot be reduced to a single 1+1 Dirac particle. In general, the complex-to-real map in arbitrary dimensions transforms a Majorana equation into a higher dimensional Dirac equation~\cite{Comment2}. Since Eq.~(\ref{eq:majorspin31}) is a Hamiltonian equation, it can be simulated in a quantum system.

The mapping of wavefunctions into larger spinors also allows us to explore exotic symmetries and unphysical operations, otherwise impossible in nature. From Eqs.~(\ref{1+1}), (\ref{map 1+1}), and (\ref{eq:majorspin31}), for the 1+1 dimensional case, we can deduce that charge conjugation is implemented in the enlarged space via the unitary operation $V_{\mathcal C}$
\begin{equation}
\label{chargeconjugation}
\psi_c  = {\mathcal C} \psi = {\mathcal U_C K} \psi \to V_{\mathcal C} \Psi = -(\sigma_z\otimes \sigma_x)\Psi.
\end{equation}
We can do something similar with time reversal, defined as the change $t\to(-t).$ In this case, we expect~\cite{Zee:2003} $
i\hbar\partial_\tau\psi'(\tau) = H\psi'(\tau) ,$ where the time variable $\tau = -t$ and the modified spinor $\psi'(\tau)=\mathcal{T}\psi(t).$ In order to preserve scalar products and distances, the time reversal operator must be an anti-unitary operator and thus decomposable as the product ${\cal T} = {\mathcal U}_{\mathcal T}\mathcal{K}.$ In $1+1$ dimensions, imposing that the Hamiltonian be invariant under time reversal, $\mathcal{T}^{-1}H\mathcal{T},$ implies that the unitary satisfies ${\mathcal U}_{\mathcal T}^{-1}(i\sigma_x\partial_x) {\mathcal U}_{\mathcal T} =-i\sigma_x\partial_x,$ with a possible choice being ${\mathcal U}_{\mathcal T} =\sigma_z.$ In other words, in the enlarged simulation space
\begin{equation}
\label{timereversal}
 \mathcal{T}\psi = {\mathcal U}_{\mathcal{T}}\mathcal{K}\psi \to V_{\mathcal T} \Psi = (\sigma_z\otimes\sigma_z)\Psi.
\end{equation}
See Fig.~\ref{fig:mapping} for a scheme of the simulated symmetries. As mentioned before, quantum simulations of unphysical operations can be straightforwardly extended to higher dimensions. In this sense, Eqs.~(\ref{chargeconjugation}) and (\ref{timereversal}) will be valid for wave functions $\psi$ of dimension $d$ as long as we consider the complex conjugation of an arbitrary wavefunction as $V_{\mathcal K} \Psi = (\sigma_z \otimes \openone_d) \Psi$.

The proposed protocol for implementing unphysical operations onto a physical setup allows us to deal with situations that are, otherwise, intractable wih conventional quantum simulations. To exemplify the value of this novel building block in the quantum simulation toolbox, we consider the case of an advanced experimental quantum simulation, impossible to reproduce with classical computers. We assume that, after a certain evolution time, it is crucial to realize an unphysical operation such as charge conjugate or time reversal, before continuing the unitary (physical) evolution. With existing tools in quantum simulations, we would need to stop the dynamics, implement a full quantum tomography of the current quantum state associated to a huge Hilbert space, apply the unphysical operation in a classical computer, encode back the modified quantum state into the experimental setup, and then to go ahead with the quantum simulation. Clearly, this task would be impossible with classical resources and would become possible with a suitable implementation of our proposed ideas.

\begin{figure*}[t]
  \centering
  \resizebox{\linewidth}{!}{\includegraphics{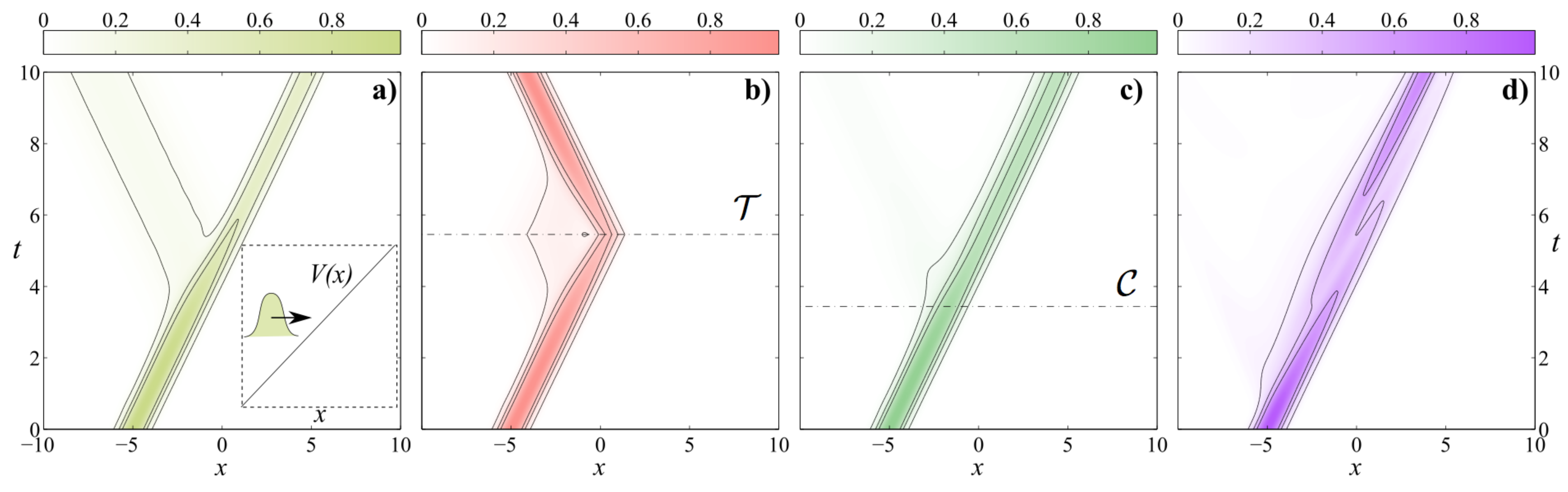}}
  \caption{Scattering of a fermion against a linearly growing potential (inset). (a) Ordinary Klein process (b) At an instant of time we apply the time reversal operator $\mathcal{T}$ causing the particle to retrace its own trajectory. (c) Similar to (b) but now we apply charge conjugation, converting the particle in its antiparticle. (d) Scattering of a Majorana particle, which propagates through the potential. Parameters are $m=0.5, c=1$ and $V(x)=x,$ in dimensionless units.\vspace{-0.5cm}}
  \label{fig:scattering}
\end{figure*}

In a recent experiment, the dynamics of a free Dirac particle was simulated with a single trapped ion~\cite{Gerritsma:2010}. Here, Eq.~(\ref{eq:majorspin31}) has a more complex structure requiring a different setup. Moreover, the encoded Majorana dynamics requires a systematic decoding via a suitable reverse mapping of observables. We can simulate Eq.~(\ref{eq:majorspin31}) in two trapped ions, with lasers coupling their internal states and motional degrees of freedom. The kinetic part, $cp_x (\openone \otimes \sigma_x)$, is created with a laser tuned to the blue and red motional sidebands of an electronic transition~\cite{Lamata:2007, Casanova:2010}, focussed on ion $2$. The second term, $\sigma_x\otimes \sigma_y$, is derived from detuned red and blue sideband excitations acting on each ion. The Hamiltonian describing this situation reads
\begin{eqnarray}
H&=&\hbar\frac{\omega_0}{2}\sigma_1^z+\hbar\frac{\omega_0}{2}\sigma_2^z+\hbar\nu a^\dag a +\hbar\nu_r b^\dag b \nonumber \\
&+& \hbar\Omega           \left[ (e^{i(qz_1 -\omega_1t+\phi_1)}+e^{i(qz_1 - \omega_{1}'t+\phi'_1)})  \sigma_1^+  + {\rm H. c.} \right]\nonumber\\
&+& \hbar\Omega           \left[ (e^{i(qz_2 - \omega_2t+\phi_2)}+e^{i(qz_2 - \omega_2't+\phi'_2)})  \sigma_2^+  + {\rm H. c.} \right]\nonumber\\
&+& \hbar{\tilde\Omega} \left[ (e^{i(qz_2-\omega t +\phi_{}  )} + e^{i(qz_2-\omega' t +\phi'_{} )})  \sigma_2^+ + {\rm H.c.}\right] . \nonumber
\end{eqnarray}
Here $z_{1,2} = Z \pm \frac{z}{2}$ are the ion positions, measured from the center of mass, $Z,$ and relative coordinate, $z$. The phases of the lasers $\phi_i$ for $i=1,2$, ($\phi$,  $\phi'$), are controlled to perform the interaction term (kinetic term). The frequencies of the center of mass and stretch mode are given by $\nu$ and $\nu_r=\sqrt{3}\nu$, while $a^{\dag}, \ a$, $b^{\dag}$, and $b$, are the corresponding creation and annihilation operators. Finally, $\Omega$ and ${\tilde \Omega}$ are the laser Rabi frequencies in the rotating-wave approximation. With the adequate choice of parameters,
\begin{equation}
\label{frecuencias}
\begin{array}{ccc}
\omega_1&=&\omega_0 + \nu_r -\delta\\
\omega_1'&=&\omega_0-\nu_r + \delta\\
\omega_2&=&\omega_0-\nu_r + \delta\\
\omega_2'&=&\omega_0 + \nu_r -\delta ,\\
\end{array} \
\begin{array}{lcc}
\omega&=&\omega_0-\nu\\
\omega'&=&\omega_0+\nu\\
\phi&=&\pi\\
\phi'&=&0,\\
\end{array} \
\begin{array}{ccl}
\phi_1&=&\pi / 2\\
\phi'_1&=&\pi /2\\
\phi_2&=&0\\
\phi'_2&=&0,
\end{array}
\end{equation}
the Hamitonian in the interaction picture reads
\begin{eqnarray}
  H&=& \hbar\eta_r\Omega(\sigma_x\otimes\openone - \openone\otimes\sigma_y ) (b^{\dag}e^{i\delta t} + be^{-i\delta t}) , \nonumber
 \\
  &&+\hbar\eta{\tilde\Omega}(\openone \otimes \sigma_x)i(a^{\dag} - a)
  \end{eqnarray}
where $\eta\equiv \eta_r3^{1/4}\equiv\sqrt{{\hbar}/{4 m' \nu}} \ll 1$ is the Lamb-Dicke parameter and $m'$ the ion mass. In the limit of large detuning, we have
$\delta\gg \eta_r\Omega\sqrt{\langle b^\dag b \rangle}, \eta{\tilde\Omega}|\langle {a^\dag -a} \rangle |$ and we recover Eq.~(\ref{eq:majorspin31}) with the momentum operator $p_x=i\hbar(a^{\dag}{-}a)/2\Delta$ and the equivalences
$c = 2\eta\Delta\tilde{\Omega}$ and $mc^2=2\hbar\eta_r^2\Omega^2/\delta$ with $\Delta=\sqrt{\hbar/4 m' \nu}$. Introducing the ratio $\gamma = |mc^2 / \langle c p_x \rangle |$, with
$\gamma = \frac{2 (\eta_r\Omega/\delta)^2}{|\langle i( a^\dagger - a)\rangle|(\eta\tilde\Omega/\delta)}$, we see it is possible to tune the numerator and denominator independently so as to preserve the dispersive regime, while exploring simultaneously the range from $\gamma \simeq 0$ (ultrarelativistic limit) to $\gamma \to\infty$ (nonrelativistic limit).

A relevant feature of the Majorana equation in 3+1 dimensions is the conservation of helicity.  A reminiscent of the latter in $1+1$ dimensions is the observable called hereafter as {\it pseudo-helicity} $\Sigma=\sigma_xp_x$. This quantity is conserved in the 1+1 Majorana dynamics of Eq.~(\ref{1+1}) but not in the 1+1 Dirac equation. We will use this observable to illustrate measurements on the Majorana wavefunction.  The mapping for operators can be simplified if we are only interested in expectation values. Reconstructing the complex spinor with the non-square matrix $\psi = M \Psi$ and $M = \left(\begin{array}{lccr} \openone &i \openone \end{array}\right)$,
associated with Eqs.~(\ref{map 1+1}) and (\ref{eq:majorspin31}), we have  $\langle O \rangle_\psi = \langle\psi| O |\psi\rangle =
  \langle \Psi| M^{\dag} O \,  M |\Psi\rangle =: \langle \tilde O\rangle_\Psi.$ Therefore, to obtain the pseudo-helicity $\Sigma$, we have to measure
\begin{equation}\label{eq:pshelicity}
\tilde\Sigma = M^{\dag} \sigma_x p_x \  M =(\openone \otimes \sigma_x - \sigma_{y} \otimes \sigma_{x})\otimes p_x
\end{equation}
in the enlarged simulation space. In ion-trap experiments, we can use laser pulses to map information about the pseudo-helicity onto the internal states. The application of a state-dependent displacement operation on ion 2, $U_2=\exp(-ik(\openone\otimes\sigma_y)\otimes p_x/2)$, generated by resonant blue and red sidebands, followed by a measurement of $\openone\otimes\sigma_z$, is equivalent to measuring the observable~\cite{Gerritsma:2010} $A(k) = \cos(k\,p_x)(\openone\otimes\sigma_z)+\sin(k\,p_x)(\openone\otimes\sigma_x)$. Here, $k$ is proportional to the probe time $t_{probe}$. Note that $\left.\frac{d}{dk}\langle A(k)\rangle\right\vert_{k=0}\propto\langle(\openone\otimes\sigma_x)\otimes p_x\rangle$. Therefore, the first term in~(\ref{eq:pshelicity}) can be measured by applying a short probe pulse to the ions and measuring the initial slope of the observable $A(k)$. To measure the second term in Eq.~(\ref{eq:pshelicity}), we apply the operation $U_1=\exp(-ik(\sigma_x\otimes\openone)\otimes p_x/2)$, and measure the spin correlation $\sigma_z\otimes\sigma_x$. We have, then, $\left.\frac{d}{dk}\langle \sigma_z \otimes \sigma_x \rangle\right\vert_{k=0} =  2 \langle (\sigma_y \otimes \sigma_x)\otimes p_x\rangle$.

So far, we have presented a complete toolbox of unphysical operations, $\mathcal{C}$, $\mathcal{T}$, and $\mathcal{K}$. We can combine all these tools to study dynamical properties of the transformed wavefunctions. To exemplify the kind of experiments that become available, we have studied the scattering of wavepackets against a linearly growing potential, $V(x)=\alpha x$, with conventional numerical tools. It is known that repulsive potentials can be penetrated by Dirac particles~\cite{Thaller:1992}, due to Klein tunneling~\cite{Casanova:2010,Gerritsma:2011}. This is shown in Fig.~\ref{fig:scattering}a, where a Dirac particle splits into a fraction of a particle, that bounces back, and a large antiparticle component that penetrates the barrier. This numerical experiment has been combined with the discrete symmetries and the Majorana equation. In Fig.~\ref{fig:scattering}b we apply the time reversal operation after the particle has entered the barrier: all momenta are reversed and the wavepacket is refocused, tracing back exactly its trajectory. In Fig.~\ref{fig:scattering}c we apply charge conjugation, changing the sign of the charge and turning a repulsive electric potential into an attractive one, which can be easily penetrated by the antiparticle. In Fig.~\ref{fig:scattering}d, we show the scattering of a Majorana particle. While there are no plane wave solutions in the Majorana equation, we can still see a wavepacket penetrating the barrier, showing a counter-intuitive insensitivity to the presence of it.

The previous implementation of discrete symmetries is valid both for Majorana and Dirac equations. Equally interesting is the possibility of combining both Dirac and Majorana mass terms in the same equation~\cite{Aste:2010}, $i \hbar \slpar \psi = m_M c \psi_{c} + m_D c \psi,$ which still requires only two ions. It also becomes feasible to have CP violating phases in the Dirac mass term, $m_D\exp(i\theta\gamma^5)$. Furthermore, we could study the dynamics of coupled Majorana neutrinos with a term $\bar{M}\psi_c$, where $\bar M$ is now a matrix and $\psi =\psi(x_1,x_2)$ is the combination of two such particles, simulated with three ions and two vibrational modes.

In summary, we have introduced a general method to implement quantum simulations of unphysical operations and non-Hamiltonian dynamics, such as the Majorana equation, in a Hamiltonian system.

\vspace*{0cm}

The authors acknowledge funding from Basque Government  grants BFI08.211, IT559-10, and IT472-10; Spanish MICINN FIS2008-05705, FIS2009-10061, and FIS2009-12773-C02-01; QUITEMAD; EC Marie-Curie program; CCQED and SOLID European projects.


\begin{thebibliography}{30}
\bibitem{Comment2}
Majorana equations in $2+1$ and $3+1$ dimensions are transformed into $3+1$ and $7+1$ Dirac equations.
\end{thebibliography}
\end{document}